\documentclass[aps,prc,twocolumn,showpacs,floatfix,nofootinbib,preprintnumbers,superscriptaddress,amsmath,amssymb]{revtex4-1}

\usepackage{epsfig}
\usepackage{graphicx}
\usepackage{stmaryrd}
\usepackage{amssymb,tabularx,dcolumn}
\usepackage{supertabular,ltxtable}
\usepackage{longtable}
\usepackage{amsmath}
\usepackage{float}
\usepackage{color}
\usepackage{bm}
\usepackage{CJKutf8}

\renewcommand{\vec}[1]{\mbox{\boldmath $#1$}}

\begin{document}

\begin{CJK*}{UTF8}{gbsn}
\title{Structure and decays of  nuclear three-body systems: the Gamow coupled-channel method in Jacobi coordinates}

\author{S.M. Wang (王思敏)}
\affiliation{FRIB/NSCL Laboratory, Michigan State University, East Lansing, Michigan 48824, USA}
\author{N. Michel}
\affiliation{FRIB Laboratory, Michigan State University, East Lansing, Michigan 48824, USA}
\author{W. Nazarewicz}
\affiliation{Department of Physics and Astronomy and FRIB Laboratory, Michigan State University, East Lansing, Michigan 48824, USA}
\author{F.R. Xu (许甫荣)}
\affiliation{School of Physics, Peking University, Beijing 100871, China}

\date{\today}

\begin{abstract}
\begin{description}
\item[Background]
Weakly bound and unbound
nuclear states appearing around particle thresholds are prototypical open quantum systems. Theories of such states must take into account  configuration mixing effects in the presence of strong coupling to the particle continuum space.
\item[Purpose]
To describe structure and decays of three-body systems, we developed a Gamow coupled-channel (GCC) approach in Jacobi coordinates by employing the complex-momentum formalism. We benchmarked the new framework  against the complex-energy Gamow Shell Model (GSM).
\item[Methods]
The GCC formalism is expressed in Jacobi coordinates, so that the center-of-mass motion is automatically eliminated. To solve 
the coupled-channel equations, we use hyperspherical harmonics to describe the angular wave functions while the radial wave functions   are expanded  in the Berggren ensemble, which includes bound, scattering and Gamow states. 
\item[Results]
We show that the GCC method is both accurate and robust. 
Its results for energies, decay widths, and nucleon-nucleon angular correlations are in good agreement with the GSM results. 
\item[Conclusions]
We have demonstrated that a three-body GSM formalism explicitly constructed in
cluster-orbital shell model coordinates provides similar results to a GCC framework expressed in  Jacobi coordinates, provided that a large configuration space is employed. 
Our calculations for $A=6$ systems and  $^{26}$O  show that nucleon-nucleon angular correlations are 
sensitive to the valence-neutron interaction. 
The new GCC technique has many attractive features
when applied to bound and unbound states of three-body systems: it is precise, efficient, and can  be extended  by introducing a microscopic model of the core.
\end{description}
\end{abstract}

\maketitle
\end{CJK*}

\section{Introduction}

Properties of rare isotopes that inhabit remote regions of the nuclear landscape at and beyond the particle driplines are in the forefront of nuclear structure and reaction research \cite{Transactions,RISAC,Dob07,For13,Bal14,NSACLRP2015}.
The next-generation of rare isotope beam  facilities will provide unique data on dripline systems that  will test  theory, highlight shortcomings,  and  identify areas for improvement. 
The challenge for nuclear  theory is to develop methodologies to reliably calculate and understand the properties and dynamics of new physical systems  with different properties due to large neutron-to-proton asymmetries  and low-lying reaction thresholds. 
Here, dripline systems are of particular interest as they can exhibit exotic radioactive  decay modes such as two-nucleon emission~\cite{Pfutzner12,Pfutzner13,Thoennessen04,Blank08,Grigorenko11,Olsen13,Kohley2013}. Theories of such nuclei must take into account their open quantum nature.
 
Theoretically, a powerful suite of $A$-body
approaches based on inter-nucleon interactions provides a quantitative description of light and medium-mass nuclei and  their reactions \cite{Elhatisari15,Navr16,Kumar17}.
To unify nuclear bound states with resonances and scattering continuum within
one consistent framework, advanced   continuum shell-model approaches have been introduced \cite{Michel09,Hagen12,Papadimitriou13}. Microscopic models of exotic nuclear states have been supplemented by a suite of powerful, albeit more phenomenological  models,  based on effective degrees of freedom such as cluster structures. While such models provide a  ``lower resolution" picture of the nucleus, they can be extremely useful when interpreting experimental data, providing guidance for future measurements, and provide guidance for more microscopic approaches.

The objective of  this work is to develop a new three-body method to describe both reaction and structure aspects of two-particle emission. A prototype system of interest  is
the two-neutron-unbound ground state of $^{26}$O
\cite{Lunderberg2012,Kohley2013,Kondo2016}. According to theory,  $^{26}$O 
 exhibits the dineutron-type  correlations
\cite{Grigorenko2015,Kondo2016,Hagino2016,Hagino16a,Fossez2017}. To describe such a system, nuclear model should be based on a fine-tuned interaction capable of describing particle-emission thresholds, a sound many-body method, and a capability to treat simultaneously bound and unbound states.

If one considers bound three-body systems, few-body models are very useful~\cite{Braaten06}, especially models based  on the Lagrange-mesh technique~\cite{Baye1994} or cluster-orbital shell model (COSM)~\cite{Suzuki1988}. However, for the description of  resonances, the outgoing wave function in the asymptotic region need to be treated very carefully. For example, one can divide the coordinate space into internal and asymptotic regions, where the R-matrix theory~\cite{Descouvemont2006,Lovell17},
microscopic cluster model \cite{Damman09}, and the diagonalization of the Coulomb interaction~\cite{Grigorenko2009} can be used. Other useful techniques include the Green function method~\cite{Hagino2016} and the complex scaling  \cite{Aoyama06,Kruppa2014}. 

Our strategy is to construct a precise three-body framework to  weakly bound and unbound systems
similar to that of the GSM~\cite{Michel2002}. The attractive feature of the GSM is that -- by employing the Berggren ensemble \cite{Berggren1968} -- it treats bound, scattering, and outgoing Gamow states on the same footing. Consequently,  energies and decay widths are obtained simultaneously as the real and imaginary parts of the complex eigenenergies of the shell model Hamiltonian \cite{Michel09}. In this study, we develop a three-body Gamow coupled-channel (GCC) approach in Jacobi coordinates with the Berggren basis. Since the Jacobi coordinates allow for the exact treatment
of  nuclear wave functions in both nuclear and asymptotic regions, and as the Berggren basis explicitly takes into account continuum effects, a comprehensive description of weakly-bound three-body systems can be achieved. As the GSM is based on the  COSM coordinates, 
a recoil term appears due to the center-of-mass motion. 
Hence, it is of interest to compare Jacobi- and COSM-based frameworks for the description of weakly bound and resonant nuclear states. 

This article is organized as follows. 
Section~\ref{model}  contains the description of  models and  approximations. In particular, it lays out the new GCC approach and GSM model used for benchmarking, and defines the configuration spaces used. 
The results for $A=6$ systems and  $^{26}$O are contained in Sec.~\ref{results}. Finally, the summary and outlook are given in Sec.~\ref{summary}.

\section{The Model}\label{model}

\subsection{Gamow Coupled Channel approach}

In the three-body GCC model, the nucleus is described in terms of a core and two valence nucleons (or clusters). The GCC Hamiltonian  can be written as:
\begin{equation}
    \hat{H} = \sum^3_{i=1}\frac{ \hat{\vec{p}}^2_i}{2 m_i} +\sum^3_{i>j=1} V_{ij}(\vec{r}_{ij})-\hat{ T}_{\rm c.m.},
\end{equation}
where $V_{ij}$ is the interaction between clusters $i$ and $j$, including central, spin-orbit and Coulomb terms, and  $\hat{T}_{\rm c.m.}$ stands for  the kinetic energy of the center-of-mass.

The unwanted feature  of three-body models is the appearance of Pauli forbidden states arising from the lack of antisymmetrization between core and valence particles.
In order to eliminate the Pauli forbidden states, we implemented the orthogonal projection method \cite{Saito69,Kuk78,Descouvemont2003} by adding to the GCC Hamiltionan the Pauli operator
\begin{equation}
\label{Pauli}
 \hat{Q}= \Lambda \sum_c |\varphi ^{j_c m_c} \rangle \langle \varphi ^{j_c m_c}|,
\end{equation}
where $\Lambda$ is a constant and $| \varphi^{j_c m_c} \rangle$ is a 2-body state involving forbidden s.p. states of core nucleons. At large values of $\Lambda$,
Pauli-forbidden states appear at high  energies, so that they are effectively suppressed.

\begin{figure}[htb]
	\includegraphics[width=0.3\textwidth]{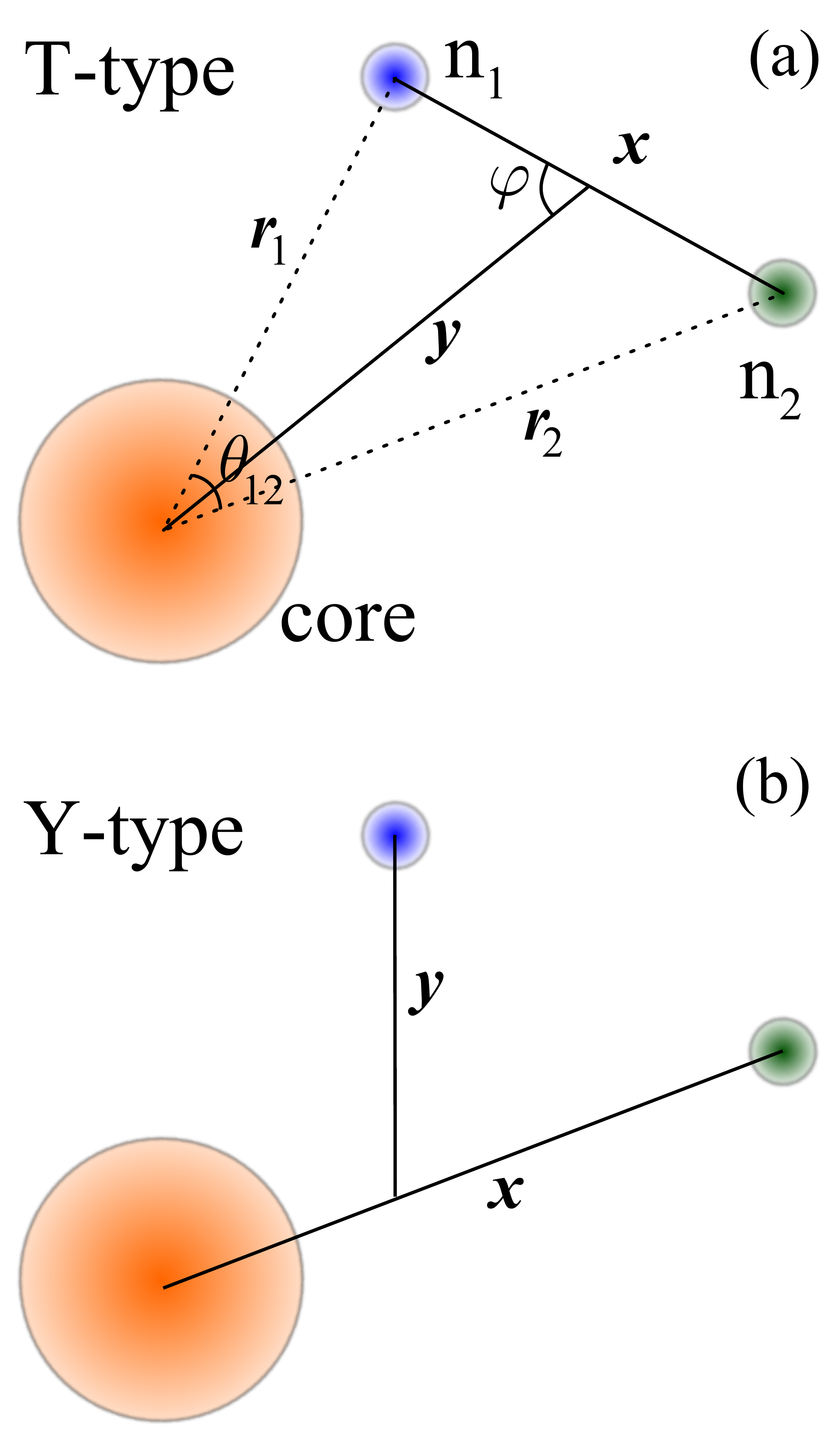}
	\caption{\label{Jacobi} Jacobi coordinates in a three-body system.}
\end{figure}

 In order to describe three-body asymptotics and to eliminate the spurious center-of-mass motion exactly, we express  the GCC model in the relative  (Jacobi) coordinates \cite{Nav00,Descouvemont2003,Navr16,Lovell17}:
\begin{equation}
\begin{aligned}
    \vec{x} &= \sqrt{\mu _{ij}} (\vec{r}_i - \vec{r}_j),\\
    \vec{y} &= \sqrt{\mu _{(ij)k}} \left(\vec{r}_k - \frac{A_i\vec{r}_i + A_j\vec{r}_j}{A_i + A_j}\right),\\
\end{aligned}
\end{equation}
where {$\vec{r}_i$} is the position vector of the i-th cluster, $A_i$ is  the i-th cluster mass number, and $\mu _{ij}$ and $\mu _{(ij)k}$ are the reduced masses associated with $\vec{x}$ and $\vec{y}$, respectively:
\begin{equation}
\begin{aligned}
    \mu _{ij} &= \frac{A_iA_j}{A_i+A_j},\\
    \mu _{(ij)k} &= \frac{(A_i+A_j)A_k}{A_i+A_j+A_k}.\\
\end{aligned}
\end{equation}
As one can see in Fig.~\ref{Jacobi}, Jacobi coordinates can be expressed as  T- and Y-types, each associated with  a complete basis set. In practice, it is convenient to calculate the matrix elements of the two-body interaction individually in T- and Y-type coordinates, and then transform them to one single Jacobi set. To describe the transformation between different types of Jacobi coordinates, it is convenient to introduce
the basis of hyperspherical harmonics (HH) \cite{Ripelle83,Kievsky08}.
The hyperspherical coordinates  are constructed from a five-dimensional hyperangular coordinates $\Omega_{5}$ and a hyperradial coordinate $\rho=\sqrt{x^2 + y^2}$.
The transformation between different sets of Jacobi coordinates is given by the Raynal-Revai coefficients~\cite{Raynal1970}.

Expressed in HH, the total wave-function can be written as \cite{Descouvemont2003}:
  \begin{equation}
    \Psi ^{JM\pi} (\rho, \Omega_5) = \rho ^{-5/2} \sum_{\gamma K} \psi ^{J\pi}_{\gamma K}(\rho) \mathcal {Y} ^{JM}_{\gamma K} (\Omega_5),
\end{equation}
where $K$ is the hyperspherical quantum number and $\gamma = \{s_1,s_2,s_3,S_{12},S,\ell_x,\ell_y,L\}$ is a set of quantum numbers other than $K$. The quantum numbers $s$ and $\ell$ stand for spin and orbital angular momentum, respectively, $\psi ^{J\pi}_{\gamma K}(\rho)$ is the hyperradial wave function, and $\mathcal {Y} ^{JM}_{\gamma K} (\Omega_5)$ is the hyperspherical harmonic. 

The resulting  Schr\"{o}dinger equation for the hyperradial wave functions can be written as a set of coupled-channel equations:\begin{widetext}
  \begin{equation}
  \label{CC}
\begin{aligned}
   \left[ -\frac{\hbar^2}{2m}\left(\frac{d^2}{d\rho^2} - \frac{(K+3/2)(K+5/2)}{\rho^2} \right)-\tilde{E} \right] \psi ^{J\pi}_{\gamma K}(\rho)& \\
   + \sum_{K'\gamma'} V^{J\pi}_{K'\gamma', K\gamma}(\rho) \psi ^{J\pi}_{\gamma'K'}(\rho)
   &+\sum_{K'\gamma'}\int_0^\infty W_{K'\gamma', K\gamma}(\rho,\rho')\psi ^{L\pi}_{\gamma'K'}(\rho')d\rho'=0,
\end{aligned}
\end{equation}
\end{widetext}
 where
  \begin{equation}
V^{L\pi}_{K'\gamma', K\gamma}(\rho) = \langle\mathcal {Y} ^{JM}_{\gamma' K'}|  \sum^3_{i>j=1} V_{ij}(\vec{r}_{ij})| \mathcal {Y} ^{JM}_{\gamma K} \rangle
\end{equation}
and 
  \begin{equation}
 W_{K'\gamma', K\gamma}(\rho,\rho') = \langle\mathcal {Y} ^{JM}_{\gamma' K'} | \hat{Q}  |\mathcal {Y} ^{JM}_{\gamma K} \rangle
\end{equation}
is the  non-local potential generated by the Pauli projection operator (\ref{Pauli}).

In order to treat the positive-energy continuum space precisely, we use the Berggren  expansion technique for the hyperradial wave function:
\begin{equation}
\label{Berggren_exp_GCC}
  \psi ^{J\pi}_{\gamma K}(\rho) = \sum_{\rm n} C^{J\pi M}_{\gamma {\rm n} K} \mathcal{B} ^{J\pi}_{\gamma {\rm n}}(\rho),
\end{equation}
where $\mathcal{B} ^{J\pi}_{\gamma {\rm n}}(\rho)$ represents a s.p. state belonging to to the Berggren  ensemble~\cite{Berggren1968}.
The Berggren ensemble defines a basis in the complex momentum plane, which includes bound, decaying, and scattering states.
The completeness relation for the Berggren ensemble can be written as:
  \begin{equation}  	
  \begin{aligned}
\sum_{{\rm n}\in b,d} \mathcal{B}_{\rm n}(k_{\rm n},\rho)\mathcal{B}_{\rm n}(k_{\rm n},\rho^\prime)  + & \int_{L^+}\mathcal{B}(k,\rho)\mathcal{B}(k,\rho^\prime)dk  \\
& = \delta(\rho-\rho^\prime),
    	\end{aligned}
\end{equation}
where $b$ are bound states and  $d$ are  decaying resonant (or Gamow)  states lying between the real-$k$ momentum axis in the fourth quadrant of the  complex-$k$ plane, and the     $L^+$ contour representing the complex-$k$ scattering continuum.
 For numerical purposes,  ${L^+}$  has to be discretized, e.g., by adopting the Gauss-Legendre quadrature~\cite{Hagen2006a}. 
In principle, the contour ${L^+}$  can be chosen arbitrarily as long as it encompasses the resonances of interest. If the contour $L^+$ is chosen to lie along the real $k$-axis, the Berggren completeness relation reduces to the Newton completeness relation \cite{Newton82} involving bound and real-momentum scattering states.

To calculate radial matrix elements  with the Berggren basis,
we employ  the exterior complex scaling~\cite{Gyarmati1971}, where integrals are calculated along a complex radial path:
    	\begin{align}
\langle\mathcal{B}_{\rm n}|&V(\rho)|\mathcal{B}_{\rm m}\rangle =\int_0^R\mathcal{B}_{\rm n}(\rho)V(\rho)\mathcal{B}_{\rm m}(\rho)d\rho \\
&+ \int_0^{+\infty}\mathcal{B}_{\rm n}(R+\rho e^{i\theta})V(R+\rho e^{i\theta})\mathcal{B}_{\rm m}(R+\rho e^{i\theta})d\rho.\nonumber
\end{align}
For potentials that decrease as $O(1/\rho^2)$ (centrifugal potential) or faster (nuclear potential), 
$R$ should be sufficiently large to bypass all singularities and the scaling angle $\theta$ is  chosen so that the integral converges, see  Ref.~\cite{Michel2003} for details. As the Coulomb potential is not square-integrable, its matrix elements diverge when  $k_n = k_m$. A practical solution is provided by the so-called ``off-diagonal method" proposed in Ref.~\cite{Michel2011}. Basically, a small offset $\pm \delta k$ is added to the linear momenta $k_n$ and $k_m$ of involved scattering wave-functions, so that the resulting diagonal Coulomb matrix element converges.

\subsection{Gamow Shell Model}

In the GSM, expressed in  COSM coordinates, one deals with the center-of-mass motion by adding a recoil term ($\hat{\vec{p}}_1\cdot\hat{\vec{p}}_2/m_nA_{\rm core}$)~\cite{Suzuki1988,Michel2002}.
The GSM Hamiltonian is diagonalized in a basis of Slater determinants built from the
one-body Berggren ensemble. In this case, it is convenient  to deal with the Pauli principle
by eliminating spurious excitations  at a level of the s.p. basis. In practice, one just needs to construct a valence s.p.  space that does not contain the orbits occupied in the core. It is equivalent to the projection technique used in GCC wherein the Pauli operator (\ref{Pauli}) expressed   in Jacobi coordinates has a two-body character.
The treatment of the interactions is the same in  GSM and GCC. In both cases,  we use the  complex scaling method to calculate  matrix elements \cite{Michel2003} and the ``off-diagonal method" to deal with the Coulomb potential~\cite{Michel2011}. 

The  two-body  recoil term is  treated in GSM  by expanding it  in a truncated basis of harmonic oscillator (HO). The HO basis depends on the oscillator length $b$ and the number of states used in the expansion. As it was demonstrated in Refs.~\cite{Hagen2006a,GSMisospin}, GSM eigenvalues and eigenfunctions converge for a sufficient number of HO states, and the dependence of the results on $b$ is very weak.

Let us note in passing that  one has to be careful when using  arguments based on the variational principle when comparing  the performance of GSM with GCC. Indeed,  the treatment of the Pauli-forbidden states is slightly different in the two approaches. Moreover,
the recoil effect in the GSM is not removed exactly. (There is no recoil term in GCC as the center-of-mass motion is eliminated through the use of Jacobi coordinates.)

\subsection{Two-nucleon correlations}

In order to study the correlations between the two valence nucleons, we  utilize the  two-nucleon density \cite{Bertsch91,Hagino05,George11} $\rho_{nn'}(r,r',\theta) = \langle \Psi|\delta(r_1-r)\delta(r_2-r^{\prime})\delta(\theta_{12} - \theta)|\Psi\rangle$, where $r_1$, $r_2$, and $\theta_{12}$ are defined in Fig.~\ref{Jacobi}(a).
In the following, we apply the normalization convention of Ref.~\cite{George11} in which the
Jacobian $8\pi^2 r^2 r'^2 \sin\theta$ is incorporated into the definition of $\rho_{nn'}$, i.e., it does not appear explicitly.
The angular density of the two valence nucleons is obtained  by integrating
$\rho_{nn'}(r,r',\theta)$ over radial coordinates:
 \begin{equation}\label{rhonn}
\rho(\theta) = \int \rho_{nn'}(r,r^\prime,\theta) dr dr^\prime.
\end{equation}
The angular density is normalized to one: 
$\int\rho(\theta) d\theta = 1.$

While it is straightforward to calculate $\rho_{nn'}$ with COSM coordinates, the angular density cannot be calculated directly with the Jacobi T-type coordinates used to diagonalize the GCC Hamiltonian.
Consequently, one can either calculate the density distribution $\rho_{\rm T}(x,y,\varphi)$ in T-type coordinates and then transform it to  $\rho(r_1,r_2,\theta_{12})$ in COSM coordinates by using the geometric relations of Fig.~\ref{Jacobi}(a),
or -- as we do in this study -- one can apply the  T-type-to-COSM 
coordinate transformation. This transformation~\cite{Raynal1970}, provides an analytical relation between hyperspherical harmonics in COSM coordinates $\mathcal {Y} ^{JM}_{\gamma^\prime K^\prime} (\vec{r}_1^\prime, \vec{r}_2^\prime )$ and the T-type Jacobi coordinates $\mathcal {Y} ^{JM}_{\gamma K} (\vec{x}^\prime, \vec{y}^\prime )$, where $\vec{r}_1^\prime$, $\vec{r}_2^\prime$, $\vec{x}^\prime$ and $\vec{y}^\prime$ are:
\begin{equation}
\begin{aligned}
\vec{r}_1^\prime &= \sqrt{A_i} \vec{r}_1,\\
\vec{r}_2^\prime &= \sqrt{A_j} \vec{r}_2,\\
\vec{x}^\prime &= \vec{x}= \sqrt{\mu_{ij}}(\vec{r}_1-\vec{r}_2),\\
\vec{y}^\prime &= \sqrt{\frac{A_i+A_j}{\mu_{(ij)k}}}\vec{y} = \frac{A_i\vec{r}_1+A_j\vec{r}_2}{\sqrt{A_i+A_j}}.
\end{aligned}
\end{equation}

\subsection{Model space and parameters}
 
In order to compare approaches  formulated in   Jacobi and COSM coordinates, we consider model spaces defined by the cutoff value $\ell_{\rm max}$, which is the maximum orbital angular momentum associated with ($\vec{r}_1$, $\vec{r}_2$) in GSM and  ($\vec{x}$, $\vec{y}$) in GCC. 
The remaining truncations come from the Berggren basis itself.

The nuclear two-body interaction between valence nucleons has been approximated by the  finite-range Minnesota force with the original parameters of Ref.~\cite{Thompson1977b}. For the core-valence Hamiltonian, we took a Woods-Saxon (WS) potential  with  parameters  fitted to the resonances of the core+$n$ system. The one- and two-body Coulomb interaction has been considered when valence protons are present.

In the case of GSM, we use the Berggren basis for the $spd$ partial waves and a HO basis for the channels with  higher orbital angular momenta. 
For $^6$He, $^6$Li and $^6$Be we assume the $^4$He core.
For $^6$He and $^6$Be, GSM we took a complex-momentum contour defined by the segments  $k= 0 \rightarrow 0.17-0.17i \rightarrow 0.34 \rightarrow 3$ (all in \,fm$^{-1}$)  for the  $p_{3/2}$ partial wave,  and 
$0 \rightarrow 0.5 \rightarrow 1 \rightarrow 3$\,fm$^{-1}$ for the remaining  $spd$ partial waves. For $^6$Li, we took the contours
$0 \rightarrow 0.18-0.17i \rightarrow 0.5 \rightarrow 3$\,fm$^{-1}$ for $p_{1/2}$; $0 \rightarrow 0.15-0.14i \rightarrow 0.5 \rightarrow 3$\,fm$^{-1}$ for $p_{3/2}$;  and $0 \rightarrow 0.25 \rightarrow 0.5 \rightarrow 3$\,fm$^{-1}$ for the $sd$ partial waves. Each segment was discretized with  10 points. This is sufficient for the energies and most of other physical quantities, but one may need more points to describe wave functions precisely, especially for the unbound resonant states that are  affected by Coulomb interaction. Hence, we choose 15 points for each segment to calculate the two-proton angular correlation of the unbound  $^6$Be. 
The  HO basis  was defined through the oscillator length $b = 2$\,fm and the maximum radial quantum number $n_{\rm max}=10$.
The WS parameters  for the $A = 6$ nuclei are: the depth of the central
term $V_{0}= 47$\,MeV;   spin-orbit strength $V_{\rm s.o.} = 30$\,MeV; diffuseness $a=0.65$\,fm; and the WS (and charge) radius  $R=2$\,fm.  With these  parameters we predict  the $3/2^-$ ground state (g.s.) of $^5$He at $E=0.732$\,MeV ($\Gamma=0.622$\,MeV), and its   first excited   $1/2^-$ state at $E=2.126$\,MeV ($\Gamma=5.838$\,MeV). 

For $^{26}$O, we consider the  $^{24}$O  core~\cite{Kanungo09,Hoffman09,Hagino2016}. In the GSM variant, we used the contour  $0 \rightarrow 0.2-0.15i \rightarrow 0.4 \rightarrow 3$\,fm$^{-1}$ for $d_{3/2}$,  and $0 \rightarrow 0.5 \rightarrow 1 \rightarrow 3$\,fm$^{-1}$ for the remaining $spd$ partial waves. For the HO basis we took $b = 1.75$\,fm and $n_{\rm max}=10$. The  WS potential for $^{26}$O  has fitted in Ref.~\cite{Hagino2016} to the resonances of $^{25}$O. Its parameters are:
$V_{0}= 44.1$\,MeV, $V_{\rm s.o.}= 45.87$\,MeV, $a= 0.73$\,fm, and $R = 3.6$\,fm.

The GCC calculations have been carried out with the maximal hyperspherical quantum number $K_{\rm max}$ = 40, which is sufficient for all the physical quantities we  study. We checked that the calculated energies  differ by as little as   2 keV when varying  $K_{\rm max}$ from 30 to 40. Similar as in  GSM, in GCC we used the Berggren basis for the $K \leqslant$ 6 channels and the HO basis for the higher angular momentum channels. The complex-momentum contour of the Berggren basis is defined as:  $k  = 0 \rightarrow 0.3-0.2i \rightarrow 0.5 \rightarrow 0.8 \rightarrow 1.2 \rightarrow 4$ (all in fm$^{-1}$), with each segment discretized with   10 points. We took the  HO basis with  $b = 2$\,fm and  $n_{\rm max} = 20$. As $k_\rho^2 = k_x^2 + k_y^2$,  the energy range covered by the GCC basis is roughly doubled as compared to that of GSM.

For the one-body Coulomb potential, we use 
the dilatation-analytic form \cite{Saito77,IdBetan08,GSMisospin}: 
\begin{equation}
U^{(Z)}_{\rm c}(r) = e^2Z_{\rm c} \frac{{\rm erf}(r/\nu_{\rm c})}{r},
\label{coul}
\end{equation}
where $\nu_c=4R_0/(3\sqrt{\pi})$\,fm,  $R_0$ is the radius of the WS potential, and $Z_{\rm c}$ is the number of core protons.

We emphasize that the large continuum space, containing states of both parities, is essential for the formation of the dineutron structure in nuclei such as $^6$He or $^{26}$O  \cite{Catara84,Pillet07,George11,Hagino14,Hagino16a,Fossez2017}. In the following, we shall study  the effect of including positive and negative parity continuum shells on the stability of threshold configurations.

\section{Results}\label{results}

\subsection{Structure of $A$=6 systems}

\begin{figure}[htb]
	\includegraphics[width=0.7\linewidth]{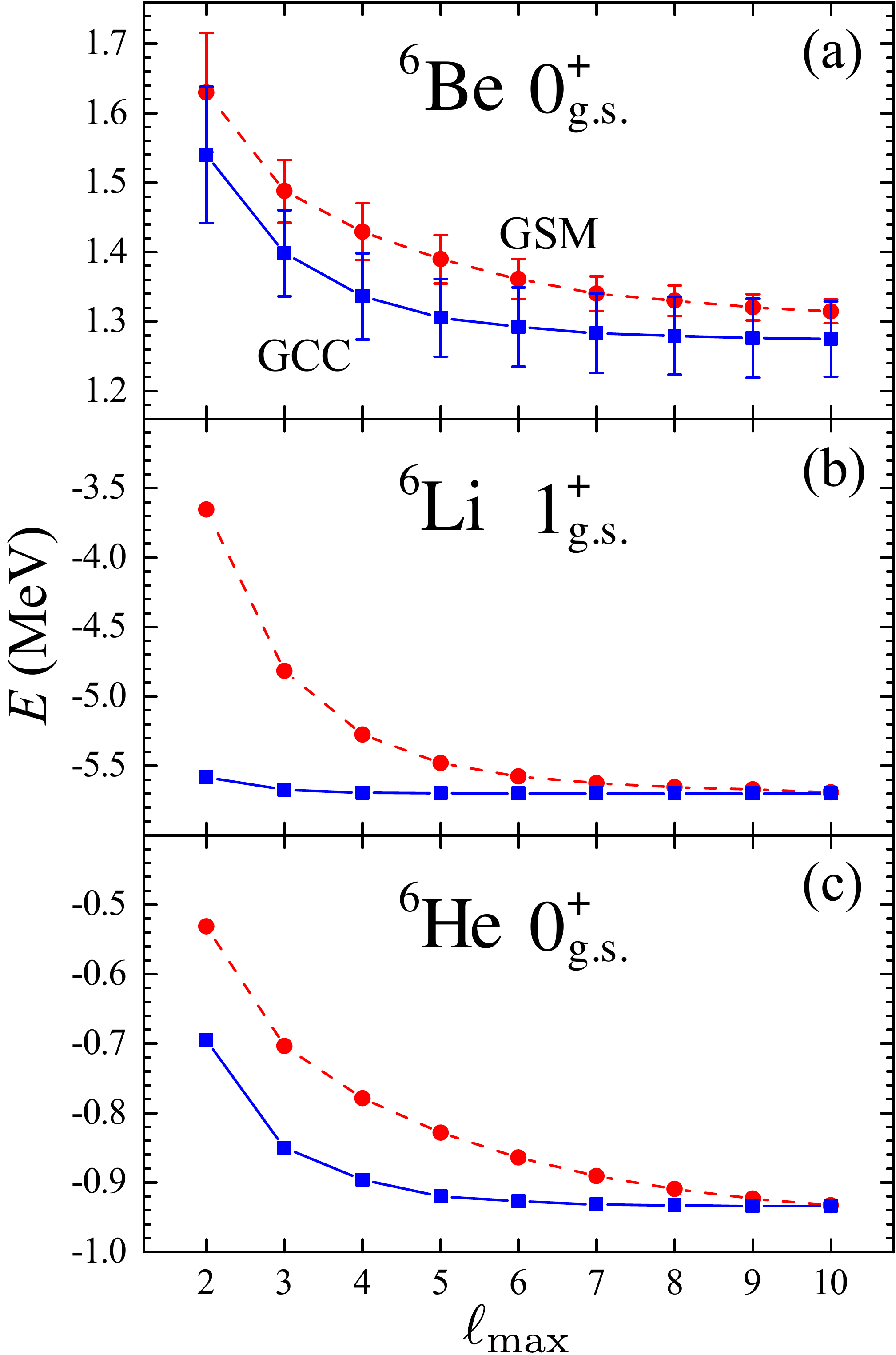}
	\caption{\label{Convergence2}
	Comparison between GSM and GCC results for 
	for the two-nucleon separation  energies of  $^6$Be,  $^6$Li, and   $^6$He obtained in different model spaces defined by  $\ell_{\rm max}$. The  bars in the panel (a) represent  decay widths.}
\end{figure}

We begin with  the GCC-GSM benchmarking for the $A=6$ systems.
Figure~\ref{Convergence2} shows the convergence rate for the g.s. energies of  $^6$He, $^6$Li, and $^6$Be with respect to $\ell_{\rm max}$. 
(See Ref.~\cite{Masui14} for a similar comparison between GSM and complex scaling results.)
While  the g.s. energies of $^6$He and $^6$Be are in a reasonable  agreement with experiment, $^6$Li is overbound. This is because the  Minnesota interaction does not explicitly
separate the $T$ = 0 and $T$ = 1 channels.  The structure of $^6$He and $^6$Be is given by the $T=1$ force, while  the $T=0$  channel that is crucial for $^6$Li has not been optimized. This is of minor importance for this study, as our goal is to benchmark GCC and GSM  not to provide quantitative predictions.  As we use different coordinates in GCC and GSM,
their  model spaces are manifestly different. Still for $\ell_{\rm max}=10$ both approaches provide very similar results, which is most encouraging.  
%%%%%%% 

One  can see in Fig. ~\ref{Convergence2} that  
the calculations done with Jacobi coordinates converge  faster than those  with COSM coordinates. This comes from the attractive character of the nucleon-nucleon interaction,  which 
results in   the presence of a di-nucleon  structure (see discussion below). Consequently, as T-type Jacobi coordinates 
well describe  the  di-nucleon cluster,  they are able to capture  correlations in a more efficient way than COSM coordinates. This is in agreement with the findings of Ref.~\cite{Kruppa2014} based on  the complex scaling  method with COSM coordinates, who obtained  the g.s. energy $^6$He  that was
slightly less bound as compared to results of Ref.~\cite{Descouvemont2003} using  Jacobi coordinates. In any case,
our calculations have  demonstrated that one obtains very similar results in GCC and GSM when  sufficiently large model spaces are considered. As shown in Table~\ref{Convergence}, the  energy difference between  GCC and GSM predictions for $A=6$ systems is very small, around 20\,keV for majority of states.
The maximum deviation of $\sim$70\,keV is obtained for the 3$^+$ state of $^6$Li. However,
because of the attractive character of the $T=0$ interaction,
the GSM calculation for this  state has not fully converged at $\ell_{\rm max} = 10$.

%%%%%
\begin{table}[!htb]
\caption{Comparison between  energies  (in MeV) and widths (in keV) 
predicted for $^6$He, $^6$Li, and $^6$Be in GSM and GCC in the $\ell_{\rm max}= 10$ model space.}
\begin{ruledtabular}
\begin{tabular}{cccc}
Nucleus  & $J^\pi$ & {GSM} & {GCC}   \\
\hline \\[-8pt]
$^6$He  & $  0^+ $  & $-0.933$  &  $-0.934$    \\
        & $  2^+ $  & ~0.800(98)  & ~0.817(42)  \\
$^6$Li  & $  1^+ $  & $-5.680$  &   $-5.698$   \\
        & $  3^+ $  & $-2.097$     &  $-2.167$   \\
        & $  0^+ $  & $-0.041$  &    $-0.048$  \\
$^6$Be  & $  0^+ $  &  ~1.314(25)  & ~1.275(54) 
\end{tabular}
\end{ruledtabular}\label{Convergence} 
\end{table}

\begin{figure}[htb]
	\includegraphics[width=0.7\linewidth]{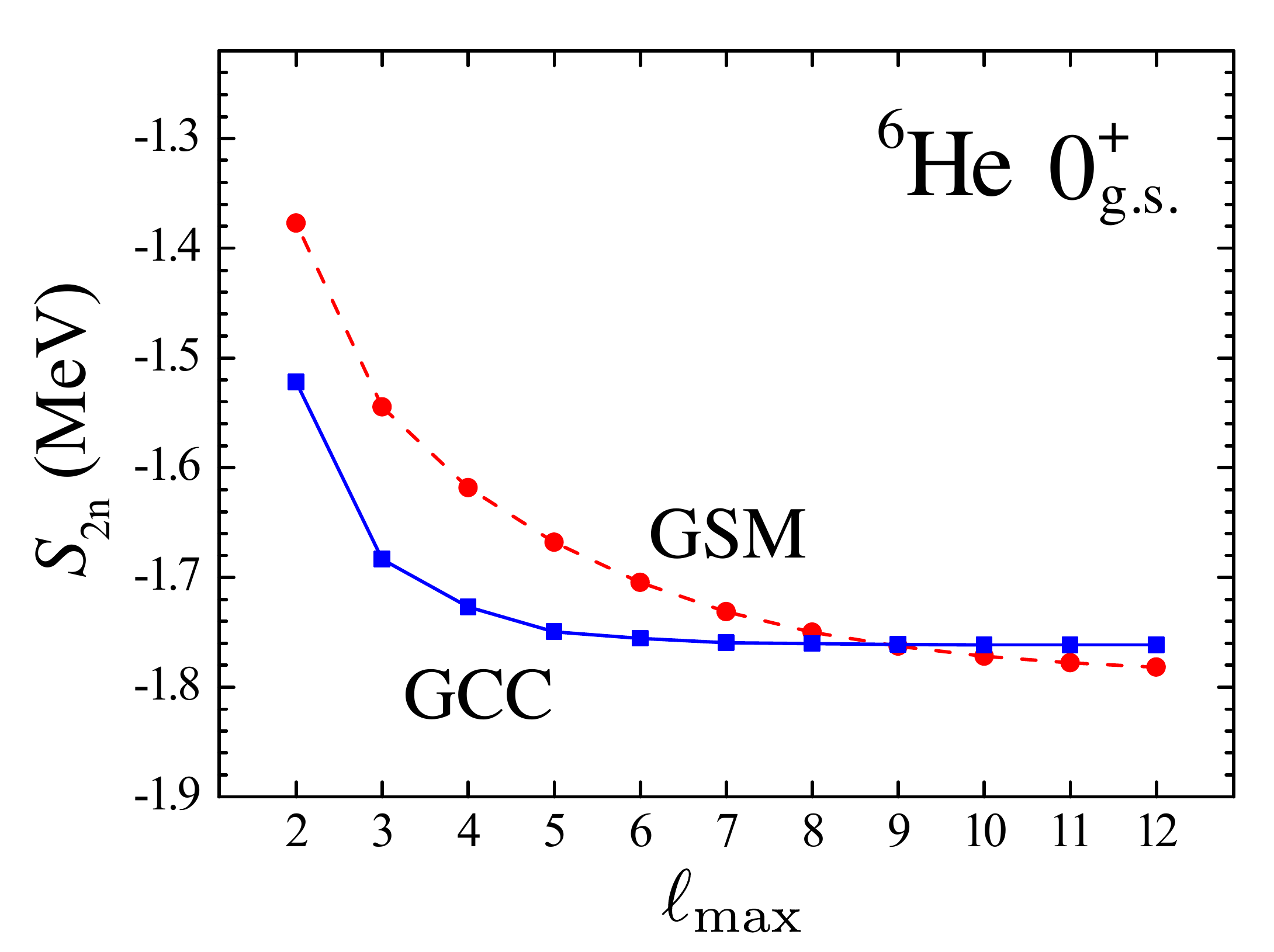}
	\caption{\label{Convergence3} Similar as in Fig.~\ref{Convergence2} but for the two-neutron separation energy of
 $^{6}$He obtained with the angular-momentum dependent Hamiltonian, see text for details.}
\end{figure}

Motivated by the discussion in Ref.~\cite{Descouvemont2003},
we have also studied the effect of   the $\ell$-dependent core-nucleus potential. To this end, we  changed the WS strength $V_0$ from 47 MeV to 49 MeV for the $\ell=1$ partial waves while keeping the standard strength for the remaining $\ell$ values. As seen in Fig.~\ref{Convergence3}, the convergence behavior obtained with  Jacobi and COSM coordinates is fairly  similar to that shown in Fig.~\ref{Convergence2}, where the WS  strength $V_0$ is the same for all partial waves. 
For $\ell_{\rm max}=12$, the difference between GSM and GCC energies of $^6$He 
becomes very small. 
This result is consistent  with the findings of  Ref.~\citep{Zhukov1993} that the 
recoil  effect can indeed 
be  successfully eliminated using COSM coordinates  at the expense of reduced convergence.
%%%%

In order to see whether the difference between the model spaces of GCC and GSM can be compensated by renormalizing
the effective  Hamiltonian,
we slightly readjusted the depth  of the WS potential in GCC calculations  to reproduce the g.s. GSM energy of $^6$He at model space $\ell_{\rm max}=7$. As a result, the strength $V_0$ changed from 47 MeV to 46.9 MeV. Except for the 2$^+$ state of $^6$He, the GSM and  GCC energies for $A=6$ systems got significantly closer as a result of such a renormalization. This indicates that the differences between Jacobi coordinates and COSM coordinates can be partly accounted for by refitting interaction parameters, even though model spaces and asymptotic behavior are different.

GCC is also in rough agreement with GSM when comparing decay widths, considering that they are very sensitive to the asymptotic behavior of the wave function, which is treated differently with Jacobi and COSM coordinates. Also, the presence of the recoil term in GSM, which is dealt with by means of the HO expansion, is expected to impact the GSM results for  decay widths.

In order to check  the precision of decay widths  calculated with GCC, we adopted the current expression \cite{Humblet}:
  \begin{equation}
  \label{width2}
  		   \Gamma  =i  \frac{\int( \Psi^{\dagger} ~\hat{\bf H}~ \Psi - \Psi ~\hat{\bf H} ~\Psi^{\dagger} )~ d{\vec{x}}d{\vec{y}}}{\int|\Psi|^2 d{\vec{x}}d{\vec{y}}},
  \end{equation}
which can be expressed in  hyperspherical coordinates as \cite{Grigorenko2000,Grig07}:
  \begin{equation}  \label{width3}
  		   \Gamma  =i \frac{\hbar^2}{m} \frac{ \left. \int d\Omega_5 {\rm Im}[\psi \frac{\partial}{\partial \rho} \psi^{\dagger}]\right|_{\rho=\rho_{\rm max}} }{\int^{\rho_{\rm max}}_0 |\psi|^2 d\rho d\Omega_5},
  \end{equation}
where $\rho_{\rm max}$  is larger than the nuclear radius (in general, the decay width should not depend on the choice of $\rho_{\rm max}$). 
By using the current expression, we obtain $\Gamma$=42\,keV for 2$^+$ state of $^6$He and $\Gamma$=54\,keV for 0$^+$ state of $^6$Be, which are practically the same as the GCC values of Table~\ref{Convergence} obtained  from the direct diagonalization.

\begin{figure}[htb]
	\includegraphics[width=0.8\linewidth]{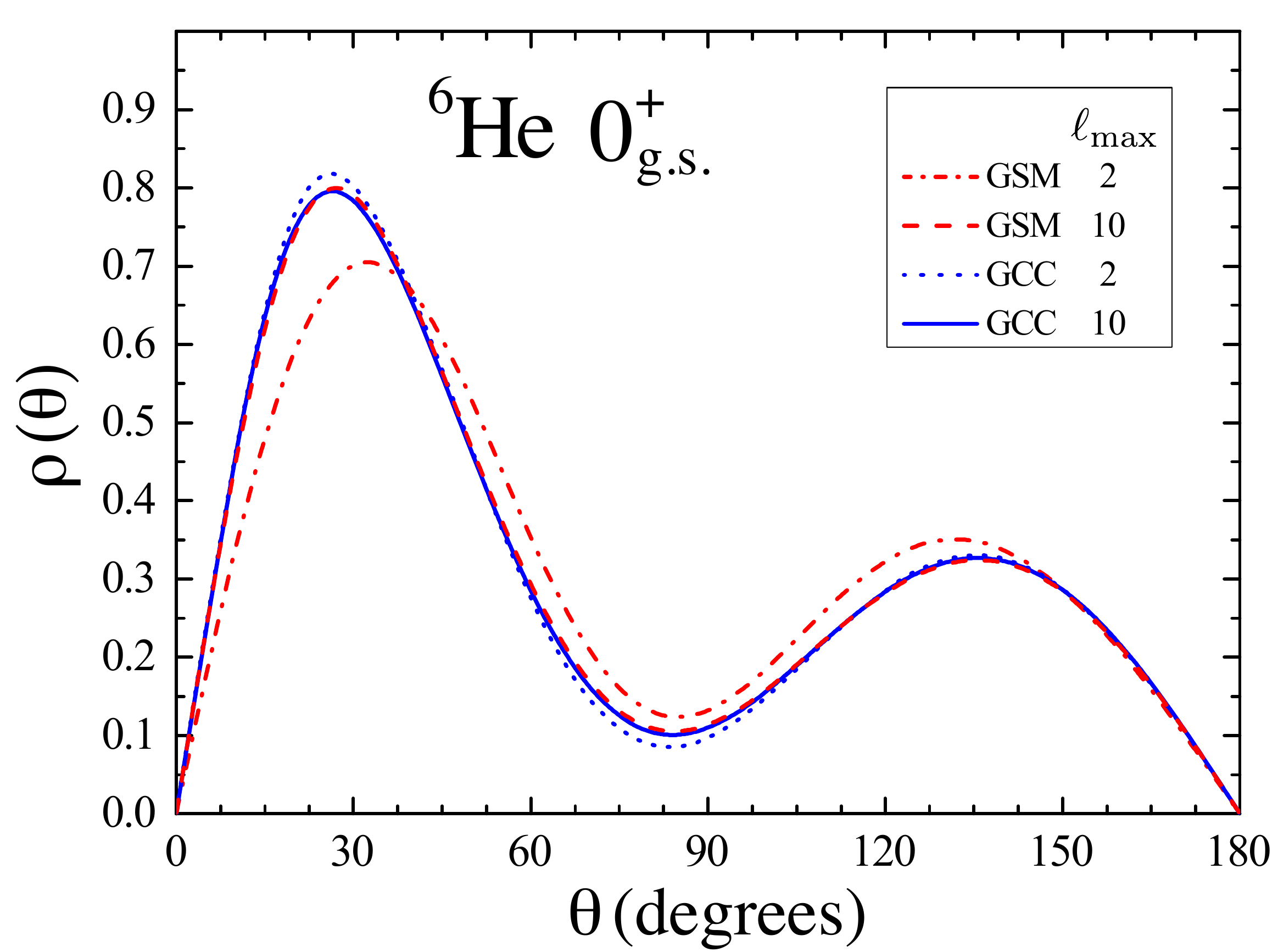}
	\caption{\label{He6_cor}Comparison between GSM and GCC results for the two-neutron angular correlation in $^{6}$He for different model spaces 
	defined by  $\ell_{\rm max}$.}
\end{figure}
%%%%

%%%%%%
\begin{figure*}[htb]
	\includegraphics[width=0.8\textwidth]{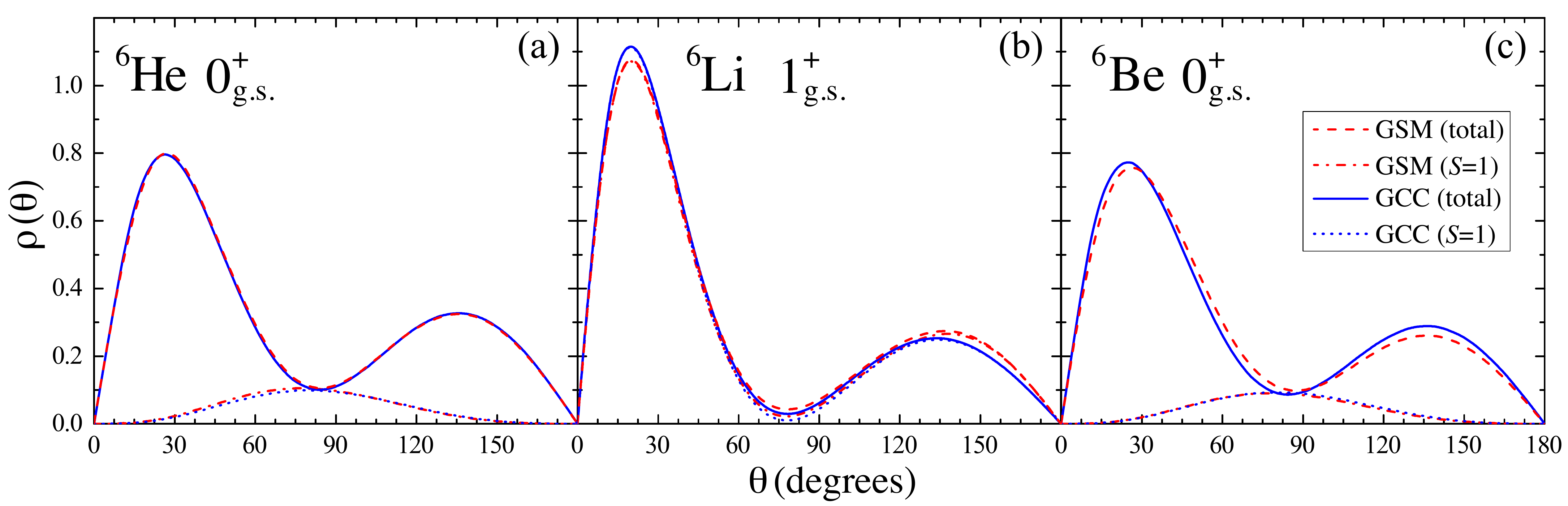}
	\caption{\label{A6_cor} Two-nucleon angular densities (total and in the $S=1$ channel)  in the g.s. configurations of $^6$He (a), $^6$Li (b), and $^6$Be (c) obtained in GSM and GCC  with $\ell_{\rm max}=10$.}
\end{figure*}
%%%%%%%

We now discuss the angular correlation  of the two valence neutrons in the g.s. of  $^{6}$He. Figure~\ref{He6_cor} shows GSM and GCC results for  model  spaces defined by  different values of $\ell_{\rm max}$. 
The distribution $\rho(\theta)$ shows two maxima \cite{Zhukov1993,Hagino05,Horiuchi07,Kikuchi10,George11,Kruppa2014,Hagino16a}. The higher peak, at a small opening angle, can be associated with a dineutron configuration. 
The second maximum, found in the region of large angles, represents the cigarlike configuration.
The GCC results for $\ell_{\rm max}=2$ and 10 are already very close. This is not the case for the GSM, which shows sensitivity to the cutoff value of $\ell$.  
This is because the large continuum space, including states of positive and negative parity is needed in the COSM picture to describe dineutron correlations \cite{Catara84,Pillet07,George11,Hagino14,Fossez2017}.
Indeed, as $\ell_{\rm max}$ increases, the angular correlations obtained in GSM and GCC are very similar. This indicates that Jacobi and COSM descriptions of $\rho(\theta)$ are  essentially equivalent  provided that the model space is sufficiently large.

In order to benchmark GCC and GSM calculations for the valence-proton case, in Fig.~\ref{A6_cor} we compare two-nucleon angular correlations for $A = 6$ nuclei $^6$He, $^6$Li, and $^6$Be. 
Similar to Refs. \cite{Hagino05,George11}, we find that the  $T=1$ configurations have a dominant 
$S = 0$ component, in which the two neutrons in $^6$He or two protons in $^6$Be are in the spin singlet state. The amplitude of the $S = 1$ density component  is  small.
For all nuclei, GCC and GSM angular correlations are close.

Similar to $^6$He, the two peaks in $^6$Be indicate diproton and cigarlike configurations~\cite{Oishi14} (see also Refs.~\cite{,Garrido07,Grigorenko09,Grig12,Egorova12,Alvarez12}).
It is to be noted  that the dineutron peak in $^6$He is slightly higher than the diproton maximum in $^6$Be. This is  due to the repulsive character of the Coulomb interaction between valence protons. 
The large maximum at small opening angles seen in $^6$Li corresponds to the deuteron-like structure. As discussed in Ref.~\cite{Horiuchi07}, this peak is larger that the the dineutron correlation in $^6$He.  Indeed, the valence proton-neutron pair in $^6$Li is very strongly correlated because the $T=0$ interaction is much stronger than the $T=1$ interaction.
The different features in the two-nucleon angular correlations  in the three  $A=6$ systems shown in Fig.~\ref{A6_cor}
demonstrate that the angular correlations  contain useful information on the effective interaction between valence nucleons.

\subsection{Structure of unbound $^{26}$O}

After benchmarking GSM and GCC for $A=6$ systems,
we apply both models to $^{26}$O, which is believed to be a threshold dineutron structure
\cite{Lunderberg2012,Kohley2013,Grigorenko2015,Kondo2016,Hagino2016,Hagino16a,Fossez2017}.
%%%%%
\begin{figure}[htb]
	\includegraphics[width=0.7\linewidth]{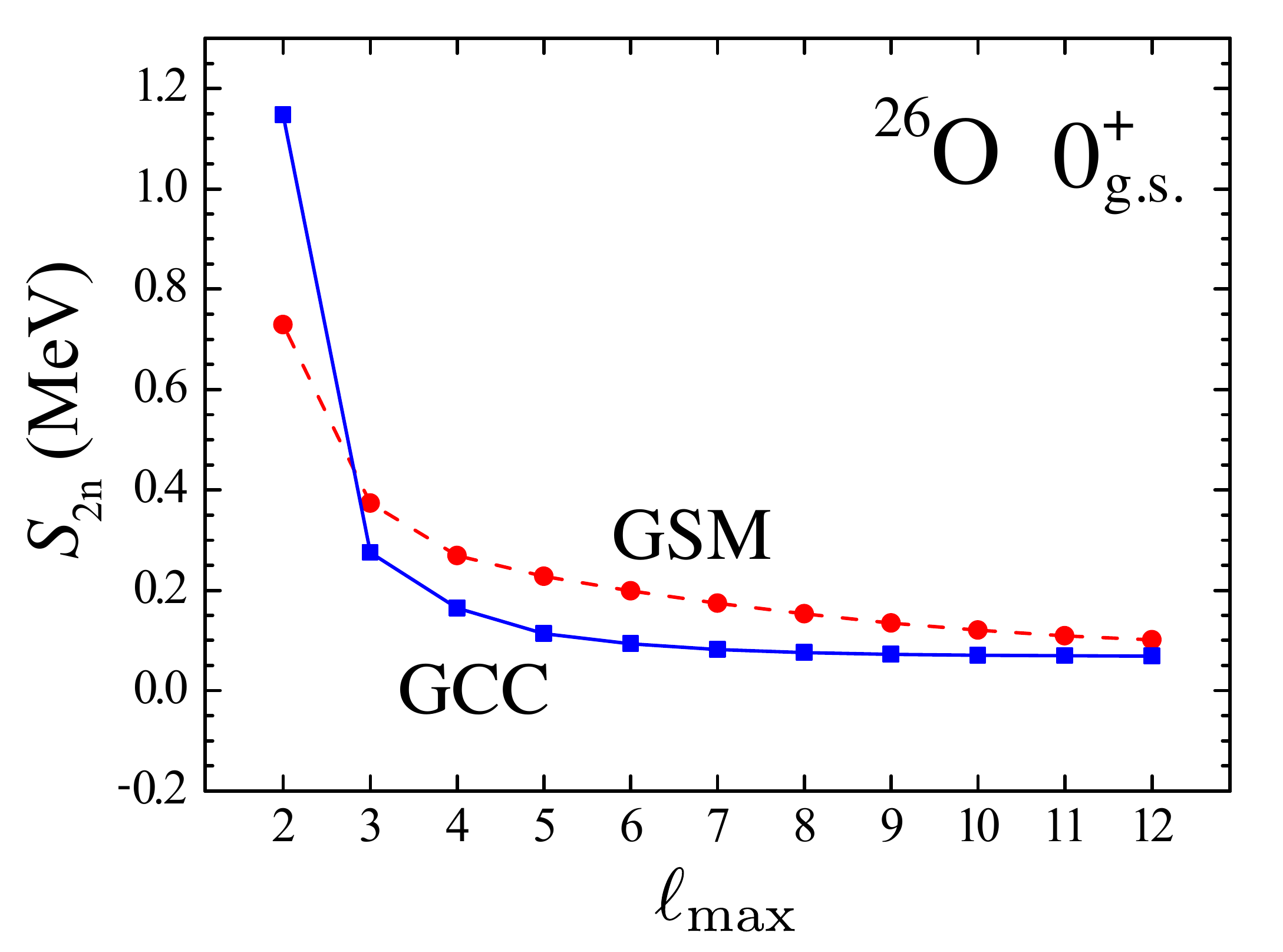}
	\caption{\label{O26_0}  Two-neutron separation energy of the g.s. of $^{26}$O computed with GSM and GCC for different values of  $\ell_{\rm max}$.}
\end{figure}
%%%%
It is a theoretical challenge to reproduce the resonances in $^{26}$O as both continuum and high partial waves must be considered.  As $^{24}$O can be 
 associated with the  subshell closure in which the $0d_{5/2}$ and $1s_{1/2}$ neutron shells
are occupied \cite{Tshoo12}, it can be used as core in our three-body model. 

Figure~\ref{O26_0} illustrates the convergence  of the g.s. of $^{26}$O with respect to  $\ell_{\rm max}$ in GSM and GCC calculations. It is seen that in the GCC approach the energy converges 
nearly exponentially  and that the stable result is practically reached at $\ell_{\rm max}=7$. 
While slightly higher in energy, the GSM results are quite satisfactory, as they differ only  by about  30\,keV from the GCC benchmark. Still, it is clear that $\ell_{\rm max}=12$ is not sufficient to reach the full convergence in GSM.

The  calculated energies and widths  of g.s. and 2$^+$ state of $^{26}$O are displayed  in Table~\ref{O26}; they are both  consistent with the most recent  experimental values~\cite{Kondo2016}. 
%%%%
\begin{table}[!htb]
\caption{Energies  and widths (all in keV)
predicted for
$^{26}$O in GSM and GCC in the $\ell_{\rm max}= 12$ model space. Also shown are the dominant 
GSM  ($\ell_1$, $\ell_2$) and GCC ($\ell_x$, $\ell_y$)  configurations.}
\begin{ruledtabular}
\begin{tabular}{ccrcr}
$J^\pi$  & \multicolumn{2}{c}{GSM} & \multicolumn{2}{c}{GCC}  \\
\hline \\[-8pt]
 $  0^+ $  & 101	  &  81\% ($d,d$) & 69 		  &  46\% ($p,p$)\\
           &          &  11\% ($f,f$) &           &  44\% ($s,s$)\\
           &          &  7\% ($p,p$)  &           &  3\%  ($d,d$) \\
 $  2^+ $  & 1137(33) &  77\% ($d,d$) & 1150(14)  &  28\% ($f,p$)\\
           &          &  7\% ($p,p$)  &           &   27\% ($p,f$)\\
           &          &   7\% ($d,s$) &           &  10\% ($d,d$)
\end{tabular}
\end{ruledtabular}\label{O26} 
\end{table}
%%%%%
The amplitudes of dominant configurations listed in  Table~\ref{O26}  illustrate the importance of considering partial waves of different parity in the GSM description of a dineutron g.s. configuration in $^{26}$O \cite{Fossez2017}.

\begin{figure}[htb]
	\includegraphics[width=\linewidth]{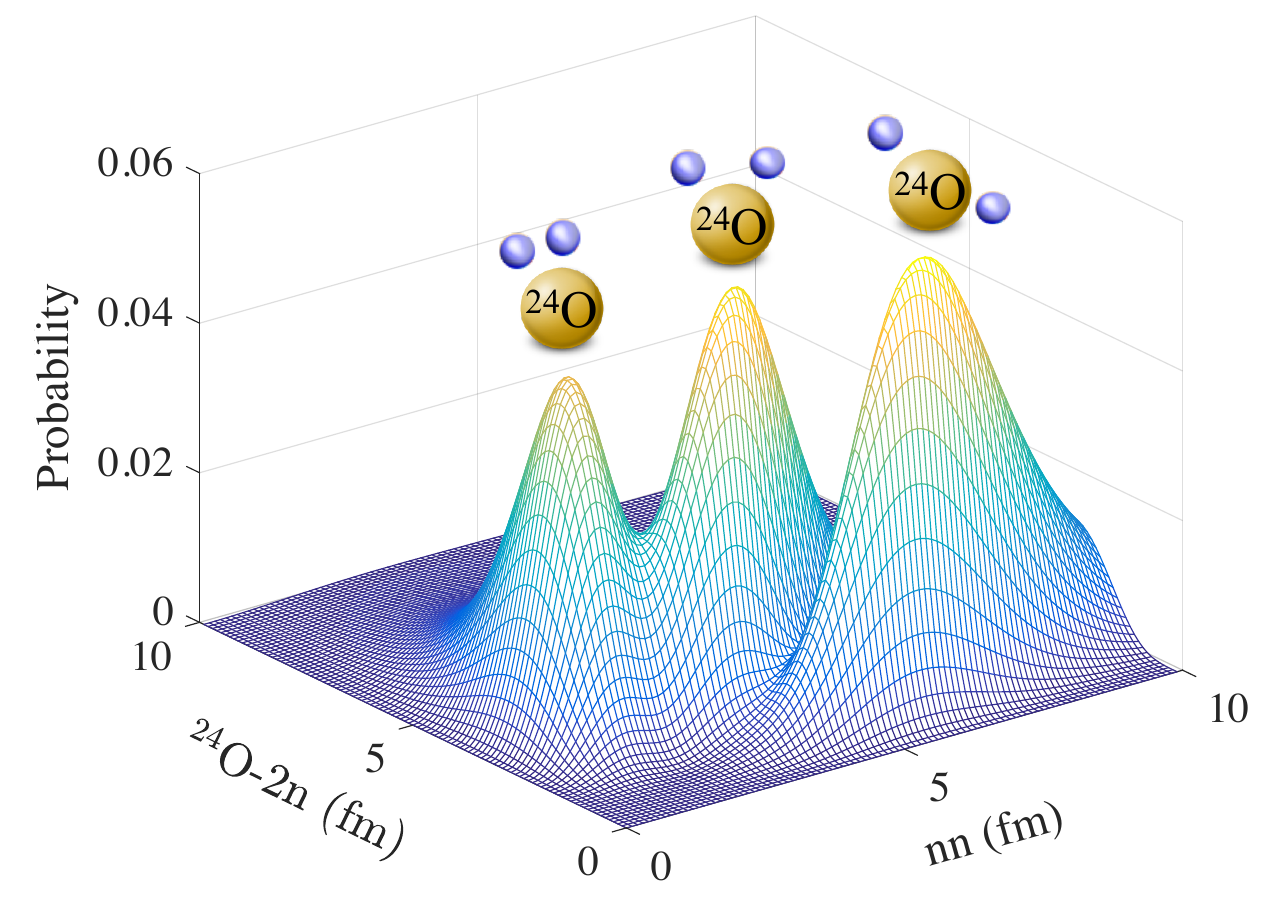}
	\caption{\label{O26_WF} GCC wave function of the g.s. of $^{26}$O in the Jacobi coordinates $nn$  and $^{24}$O$-2n$. 
	}
\end{figure}
\begin{figure}[htb]
	\includegraphics[width=0.8\linewidth]{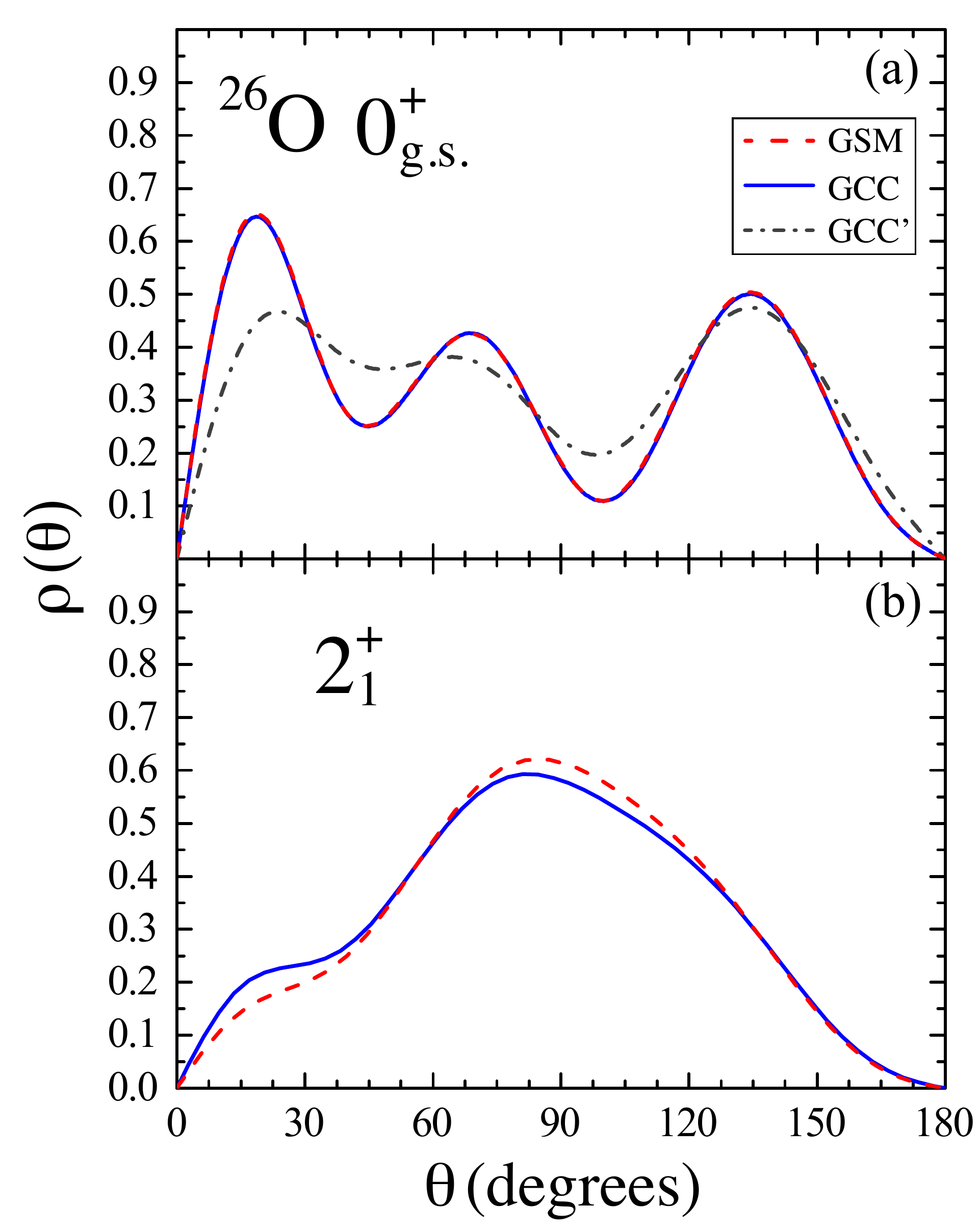}
	\caption{\label{O26_cor} Two-neutron angular correlation for the  0$^+$ g.s. (a) and 2$^+_1$  state (b) configuration of $^{26}$O computed with GCC (solid line) and GSM (dashed line) with $\ell_{\rm max}=10$.  The dash-dotted curve labeled GCC' in panel (a) shows GCC results obtained with the strength of the neutron-neutron interaction reduced by 50\%.}
\end{figure}

The g.s. wave function of $^{26}$O  computed in GCC is shown in  Fig.~\ref{O26_WF} in the Jacobi coordinates. The corresponding angular distribution is displayed in Fig.~\ref{O26_cor}.
%%%%%%
%%%
Three pronounced peaks associated with the dineutron, triangular, and cigarlike configurations~\cite{Hagino2016,Hove2017} can be identified. 
In GCC, the ($\ell_x$, $\ell_y$) = ($s, s$), ($p, p$)  components dominate the g.s. wave function of  $^{26}$O; this is consistent with a sizable clusterization of the two neutrons.
In COSM coordinates,  it is the ($\ell_1$, $\ell_2$) = ($d,d$) configuration that dominates, but the the negative-parity ($f, f$) and ($p, p$) channels contribute with $\sim$20\%.
Again, it is encouraging to see that with $\ell_{\rm max}=10$ both approaches predict very similar two-nucleon densities.

In Table~\ref{O26}  we also display the predicted structure of the excited 2$^+$ state of $^{26}$O . The predicted energy is close to experiment~\cite{Kondo2016} and other theoretical studies, see, e.g., \cite{Hagino2016,Grigorenko2015,Volya2006,Tsukiyama2015,Bogner2014}. 
We obtain a small width for this state, which is consistent with the GSM+DMRG calculations of Ref.~\cite{Fossez2017}.
The GCC occupations of Table~\ref{O26} indicate that the wave function of the 2$^+$ state is spread  out in space, as the main three configurations, of cluster type, only contribute to the wave function with only 65\%.
When considering the GSM wave function, the ($d, d$) configuration  dominates.  The  corresponding two-neutron angular correlation shown in Fig.~\ref{O26_cor}(b)  exhibits a broad distribution with a maximum  around 
90$^\circ$. 
This situation is fairly similar to what has been  predicted for the 2$^+$ state of $^6$He \cite{George11,Kruppa2014}.

Finally, it is interesting to study  how the neutron-neutron interaction impacts the angular correlation.
To this end, Fig.~\ref{O26_cor}(a) shows $\rho(\theta)$ obtained with  
the Minnesota neutron-neutron interaction whose strength has been reduced by 50\%. 
While there are still three peaks present, the distribution becomes more uniform and 
the dineutron component no longer dominates. We can this conclude that the $nn$ angular correlation
can be used as an indicator of the  interaction between valence nucleons.

\section{Conclusions}\label{summary}

We developed a Gamow coupled-channel approach in Jacobi coordinates with the Berggren basis to describe structure and decays of three-body systems. We benchmarked  the performance of the new approach against the Gamow Shell Model. Both methods are capable of considering large continuum spaces but differ in their treatment of  three-body asymptotics, center-of-mass motion, and Pauli operator.
In spite of these differences, we demonstrated that the Jacobi-coordinate-based framework (GCC) and COSM-based framework (GSM) can produce fairly similar results, provided that the continuum space is sufficiently large.

For benchmarking and illustrative examples we choose  $^6$He, $^6$Li, and $^6$Be,   and $^{26}$O -- all viewed as a core-plus-two-nucleon systems. We discussed the spectra, decay widths, and nucleon-nucleon angular correlations in these nuclei.
The Jacobi coordinates capture  cluster correlations (such as dineutron and deuteron-type) more efficiently; hence, the convergence rate of GCC is faster than that of GSM. 

For  $^{26}$O, we demonstrated the sensitivity of $nn$ angular correlation to the valence-neutron interaction. It will be interesting to investigate this aspect further to provide 
guidance for future experimental investigations of di-nucleon correlations in bound and unbound states of dripline nuclei.

In summary, we developed an efficient approach to structure and decays of three-cluster systems. The GCC method is based on a Hamiltonian involving a two-body interaction between valence nucleons and a
one-body field representing the core-nucleon potential. The advantage of the model is its ability to correctly describe the three-body asymptotic behavior and the efficient treatment of the continuum space, which is of particular importance for the treatment of threshold states and narrow resonances. The model can be easily extended along the lines of the resonating group method   by introducing a microscopic picture of the core \cite{Navr16,Jaganathen14}. Meanwhile, it can be used to elucidate  experimental findings on  dripline systems, and to provide finetuned  predictions to guide  $A$-body approaches.

\begin{acknowledgments}
We thank K{\'e}vin Fossez, Yannen Jaganathen,  Georgios Papadimitriou, and Marek P{\l}oszajczak for useful discussions. 
This material is
based upon work supported by the U.S.\ Department of Energy, Office of
Science, Office of Nuclear Physics under  award numbers 
DE-SC0013365 (Michigan State University),   DE-SC0008511 (NUCLEI
SciDAC-3 collaboration), DE-SC0009971 (CUSTIPEN: China-U.S. Theory Institute for Physics with
Exotic Nuclei),
and also supported in part by Michigan State University through computational resources provided by the Institute for Cyber-Enabled Research.
\end{acknowledgments}

\end{document}